\documentclass[prd, tightten, aps, nofootinbib, showpacs, preprintnumbers]{revtex4}
\begin{document}
\preprint{USM-TH-222}
\title{On scale symmetry breaking and confinement in $D=3$ models}
\author{Patricio Gaete}
\email{patricio.gaete@usm.cl}
\affiliation{Departamento de F\'{\i}sica, Universidad T\'{e}cnica
Federico Santa Mar\'{\i}a, Valpara\'{\i}so, Chile}
\author{Jos\'{e} A. Hela\"{y}el-Neto}
\email {helayel@cbpf.br} \affiliation{Centro Brasileiro de Pesquisas
F\'{\i}sicas, Rua Xavier Sigaud, 150, Urca, 22290-180, Rio de
Janeiro, Brazil}
\date{\today}

\begin{abstract}
Within the framework of the gauge-invariant, but path-dependent,
variables formalism, we study the connection between scale symmetry
breaking and confinement in three-dimensional gluodynamics. We
explicitly show that the static potential profile contains a linear
potential, leading to the confinement of static charges. Also, we
establish a new type of equivalence among different
three-dimensional effective theories.
\end{abstract}
\maketitle

\section{Introduction}

One of the long-standing issues in non-Abelian gauge theories is a
quantitative description of confinement. In this context, it may be
recalled that phenomenological models have been of considerable
importance in our present understanding of the physics of
confinement, and can be considered as effective theories of QCD. One
of these, which is the dual superconductivity picture of QCD
\cite{Nambu}, has probably enjoyed the greatest popularity. In this
picture, the crucial feature is the condensation of topological
defects originated from quantum fluctuations (monopoles).
Accordingly, the color electric flux linking quarks is squeezed into
strings (flux tubes), and the nonvanishing string tension represents
the proportionality constant in the linear, quark confining,
potential. Lattice calculations have confirmed this picture by
showing the formation of tubes of gluonic fields connecting colored
charges \cite{Capstick}. Recently, 't Hooft \cite{'t Hooft} has
suggested a new approach to the confinement problem which includes a
linear term in the dielectric field that appears in the energy
density. It should be highlighted at this point that QCD, at the
classical level, possesses scale invariance which is broken by
quantum effects. Interestingly, these effects can be described by
formulating classical gluodynamics in a curved space-time with non-
vanishing cosmological constant. More precisely, an effective low-
energy Lagrangian for gluodynamics which describes semi-classical
vacuum fluctuations of gluon field at large distances is obtained
\cite{Levin}, where a dilaton coupling to gauge fields plays an
essential role in this development.

On the other hand, in recent times the connection between scale
symmetry breaking and confinement in terms of the gauge-invariant
but path-dependent variables formalism, has been developed
\cite{GaeteS,GaeteS2}. In particular, for a phenomenological model
which contains both a Yang-Mills and a Born-Infeld term, we have
shown the appearance of a Cornell-like potential which satisfies the
't Hooft basic criterion, after spontaneous breaking of scale
invariance in both Abelian and non-Abelian cases \cite{GaeteS}. It
is worthy noting here that similar results have been obtained in the
context of gluodynamics in a curved space-time \cite{GaeteS3}. In
fact, we have shown the vital role played by the massive dilaton
field in triggering a linear potential, leading to the confinement
of static charges. Accordingly, this picture may be considered as
equivalent to that based on the condensation of topological defects.
In this way, we have established a new correspondence between these
two non-Abelian effective theories. The present work is aimed at
studying the stability of the above scenario for the
three-dimensional case. The main purpose here is to reexamine the
effects of the dilaton field on a physical observable, and to check
if a linearly increasing gauge potential is still present whenever
we go over into three dimensions.

Before going ahead, we would like to recall a number of motivations
to undertake our study $(2+1)$-D. In this space-time, Yang-Mills
theories are super-renormalizable and mass for the gauge fields are
not in conflict with gauge symmetry \cite{Deser}. Indeed,
topologically massive Yang-Mills theories are a very rich field of
investigation and it has been shown that Yang-Mills-Chern-Simons
models are actually ultra-violet finite \cite{turmaRJ}. Yet,
$(2+1)$-D theories may be adopted to describe the high-temperature
limit of models in $(3+1)$-D \cite{Das}. Planar gauge theories are
also of interesting to probe low-dimensional Condensed Matter
systems, such as the description of bosonic collective excitations
(like spin or pairing fluctuations) by means of effective gauge
theories and high-$T_C$ superconductivity, for which planarity is a
very good approximation \cite{Khveshchenko}. We should also mention
that $(2+1)$-D theories, specially Yang-Mills theories, are very
relevant for a reliable comparison between results coming from the
continuum and lattice calculations, for much larger lattices can be
implemented in three space-time dimensions \cite{Reinhardt}. Most
recently, $3D$ physics has been raising a great deal of interest in
connection with branes activity; in this context, issues like
self-duality \cite{Singh} and new possibilities for supersymmetry
breaking as induced by $3$-branes \cite{Kol} are of special
relevance. In addition, the study of the quark-antiquark potential
for some non-Abelian $(2+1)$-dimensional Yang-Mills theories has
been considered in \cite{Orland}.

In order to accomplish the purpose of probing different aspects of
three-dimensional field-theoretic models, we shall work out the
static potential for three-dimensional gluodynamics in curved
space-time along the lines of Ref. \cite{GaeteS,GaeteS3}. Our
treatment is manifestly gauge-invariant for the static potential.
This analysis give us an opportunity to compare our procedure with
related three-dimensional models. As will be seen, the
three-dimensional gluodynamics version is equivalent to a Lorentz-
and CPT- violating Maxwell-Chern-Simons model, while a
three-dimensional phenomenological model which includes a Yang-Mills
and a Born-Infeld term is equivalent to the above models in the
short-distance regime. One important advantage of this approach is
that it allows us to describe different models in an unified way.
The point we wish to emphasize, however, is that we once again
corroborate that confinement arises as an Abelian effect. In
general, this picture agrees qualitatively with that of Luscher
\cite{Luscher}. More recently, it has been related to relativistic
membrane dynamics in \cite{Gabadaze}, and implemented through the
Abelian projection method in \cite{Kondo}.

\section{Interaction energy}

We turn now to the problem of obtaining the interaction energy
between static point-like sources for the three models we shall
consider in this work. With this purpose, we shall compute the
expectation value of the energy operator, $H$, in the physical
state, $|\Phi\rangle$, describing the sources, which we will denote
by $ {\langle H\rangle}_\Phi$. We begin by summarizing very quickly
the dilaton effective Lagrangian coupled to gluodynamics, and
introduce some notation that is needed for our subsequent work. We
start from the four-dimensional space-time Lagrangian density
\cite{Levin}:
\begin{equation}
{\cal L}^{(3+1)} = \frac{{\left| {\varepsilon _V } \right|}}{{m^2
}}\frac{1}{2}e^{{\raise0.7ex\hbox{$\chi $} \!\mathord{\left/
 {\vphantom {\chi  2}}\right.\kern-\nulldelimiterspace}
\!\lower0.7ex\hbox{$2$}}} \left( {\partial _\mu  \chi } \right)^2  +
\left| {\varepsilon _V } \right|e^\chi  \left( {1 - \chi } \right)
- e^\chi  \left( {1 - \chi } \right)\frac{1}{4}F_{\mu \nu }^a
F^{a\mu \nu }, \label{dil5}
\end{equation}
where the real scalar field (dilaton) $\chi$, of mass $m$, describes
quantum fluctuations, and $- \left| {\varepsilon _V } \right|$ is
the vacuum energy density. Let us also mention here that the stable
minimum is in $\chi=0$, according to the work of Ref.
\cite{Shifman}. Following our earlier procedure \cite{GaeteS3}, we
shall now consider the expansion near $\chi=0$. In such a case,
expression (\ref{dil5}) becomes
\begin{equation}
{\cal L}_{eff}^{(3+1)}  =  - \frac{1}{4}F_{\mu \nu }^a \left( {1 +
\frac{{m^2 }}{\Delta _{(3+1)}}} \right)F^{a\mu \nu } + \frac{{m^2
}}{{32\left| {\varepsilon _V } \right|}}\left( {F_{\mu \nu }^a }
\right)^2 \frac{1}{\Delta _{(3+1)}}\left( {F_{\mu \nu }^a }
\right)^2 + \left| {\varepsilon _V } \right|. \label{dil10}
\end{equation}
To get the last expression, we have integrated over the $\chi$-
field. Next, in order to linearize this theory , we introduce the
auxiliary field, $\phi$. Then, we write
\begin{equation}
{\cal L}_{eff}^{(3+1)}  =  - \frac{1}{4}F_{\mu \nu }^a \left( {1 +
\frac{{m^2 }}{\Delta _{(3+1)}}} \right)F^{a\mu \nu }  +
\frac{1}{2}\left( {\partial _\mu  \phi } \right)^2 -
\frac{1}{4}\frac{m}{{\sqrt {\left| {\varepsilon _V } \right|} }}\phi
\left( {F_{\mu \nu }^a } \right)^2 + \left| {\varepsilon _V }
\right|. \label{dil15}
\end{equation}
Once again, by expanding about $\phi=\phi_0$, we then get
\begin{equation}
{\cal L}_{eff}^{(3+1)}  =  - \frac{1}{4}F_{\mu \nu }^a
\frac{1}{\varepsilon }\left( {1 + \frac{{\varepsilon m^2 }}{\Delta
_{(3+1)}}} \right)F^{a\mu \nu } + \left| {\varepsilon _V } \right|
, \label{dil20}
\end{equation}
where $\frac{1}{\varepsilon } \equiv 1 + \frac{m}{{\sqrt {\left|
{\varepsilon _V } \right|} }}\phi _0$. Notwithstanding, in order to
set-up the context for our discussion, it is useful to recall that
the field configuration $\phi_0$ must be constant, so that the terms
in ${\dot{\phi}} ^2 $ and $(\nabla \phi )^2$ do not add positive
contributions to the energy. Actually, we must have that $\phi_0$ is
zero. To understand why $\phi_0$ must be zero, we examine its
contribution to the density energy, $\Theta ^{00}$; with space-time
independent $\phi_0$, we have
\begin{equation}
\Theta ^{00}  = \frac{1}{2}\frac{m}{{\left| {\varepsilon _V }
\right|}}\phi _0 \left( {{\bf E}^a \cdot {\bf E}^a  + {\bf B}^a
\cdot {\bf B}^a } \right), \label{dil20a}
\end{equation}
where ${\bf E}^a$ and ${\bf B}^a$  are respectively the electric and
magnetic fields. To minimize such a term, we see that $\phi_0$ must
be zero and the minimum of energy turns out to be $-\left|
{\varepsilon _V } \right|$, according to what discussed in
\cite{Levin}. In such a case, the expression (\ref{dil20}) reduces
to
\begin{equation}
{\cal L}_{eff}  =  - \frac{1}{4}F_{\mu \nu }^a \left( {1 + \frac{{
m^2 }}{\Delta }} \right)F^{a\mu \nu } + \left| {\varepsilon _V }
\right| . \label{dil20b}
\end{equation}

Our immediate undertaking is to obtain the corresponding effective
Lagrangian density in $(2+1)$ dimensions. In other terms, this means
that we have to compactify one spacelike dimension. In order to do
so, we employ a sort of Kaluza-Klein approach \cite{GaeteSpallucci},
where the limit of infinite compactification radius is obtained by
means of a self-consistency condition, as we shall see below.
According to this idea, one writes
\begin{equation}
{\cal L}_{eff}^{(2 + 1)}  =  - \frac{1}{4}F_{\mu \nu }^a
\sum\limits_n  \left( {1 + \frac{{m^2 }}{{\Delta _{(2 + 1)} + a^2
}}} \right)F^{a\mu \nu } + \left| {\varepsilon _V } \right|,
\label{dil25}
\end{equation}
with $a^2  \equiv {\raise0.7ex\hbox{${n^2 }$} \!\mathord{\left/
 {\vphantom {{n^2 } {R^2 }}}\right.\kern-\nulldelimiterspace}
\!\lower0.7ex\hbox{${R^2 }$}}$ , and $R$ is the compactification
radius. We see, therefore, that the novel feature of the present
theory is the presence of the $a^{2}$-term. Such a  question
motivates us to study the role of the dilaton field in the
three-dimensional case. Having characterized the new effective
Lagrangian, we can now compute the interaction energy for a single
mode in Eq.(\ref{dil25}). To this end, we shall first examine the
Hamiltonian framework for this theory. The canonical momenta are
$\Pi ^{a\mu } = - \left( {1 + \frac{{ m^2 }}{\Delta + a^2 }}
\right)F^{a0\mu } $, which results in the usual primary constraint,
$\Pi^{a0}=0$, and $\Pi ^{ai }  =  - \left( {1 + \frac{{m^2 }}{\Delta
+ a^2 }} \right)F^{a0i }$. Here, we have simplified our notation by
setting $\Delta_{(2+1)}\equiv\Delta$. This allows us to write the
following canonical Hamiltonian:
\begin{equation}
H_C = \int {d^2 x} \left\{ {\frac{1 }{2}\Pi ^{ai} \left( {1 +
\frac{{ m^2 }}{\Delta + a^2 }} \right)^{ - 1} \Pi ^{ai}+{\frac{1}{{4
}}F_{ij}^a \left( {1 + \frac{{ m^2 }}{\Delta + a^2 }} \right)F^{aij}
}}+{\Pi ^{ai} \left( {\partial _i A_0^a + gf^{abc} A_0^c A_i^b }
\right)} \right\} . \label{dil30}
\end{equation}
The secondary constraint generated by the time preservation of the
primary constraint $\Pi^{a0}\approx0$ is now $\Gamma ^{a \left( 1
\right)} \left( x \right) \equiv \partial _i \Pi ^{ai}  + gf^{abc}
A^{bi} \Pi _i^c \approx 0$. It is straightforward to check that
there are no more constraints and that the above constraints are
first class. The corresponding extended Hamiltonian (that generates
translations in time) is given by $H = H_C + \int d x \left(
{c_0^{a} (x)\Pi_0^{a} (x) + c_1^{a} (x)\Gamma ^{a \left( 1 \right)}
\left( x \right)} \right)$, where $c_0^{a}(x)$ and $c_1^{a}(x)$ are
arbitrary multipliers. Since $\Pi^{0a} = 0$ always, and ${\dot
{A}}_0^a \left( x \right)= \left[A_0^a \left( x \right),H \right] =
c_0^{a} \left( x \right)$, the dynamical variables $ A^{0a}$ and
their conjugate $\Pi^{0a}$ may now be eliminated from the theory. We
therefore drop the term in $\Pi^{0a}$ and define a new arbitrary
coefficient $c^a  \left( x \right) = c_1^{a} \left( x \right) -
A_0^{a} \left( x \right)$. The Hamiltonian then reduces to
\begin{equation}
H = \int {d^2 x} \left\{ {\frac{1}{2}{\bf \Pi} ^{a} \left( {1 +
\frac{{ m^2 }}{\Delta + a^2 }} \right)^{ - 1} {\bf \Pi} ^{a}
}+{\frac{1}{{4 }}F_{ij}^a \left( {1 + \frac{{ m^2 }}{\Delta + a^2 }}
\right)F^{aij} }+ { c^a \left( x \right) \left( {\partial _i
\Pi^{ai} + gf^{abc} A^{bi}\Pi _i^c } \right)} \right\}.
\label{dil35}
\end{equation}
In order to break the gauge freedom of the theory, we introduce a
gauge-fixing condition such that the full set of constraints becomes
second class; so, we choose
\begin{equation}
\Gamma^{a\left( 2 \right)} \left( x \right) = \int\limits_0^1
{d\lambda } \left( {x - \xi } \right)^i A_i^{\left( a \right)}
\left( {\xi  + \lambda \left( {x - \xi } \right)} \right) \approx 0,
\label{dil40}
\end{equation}
where  $\lambda$ $(0\leq \lambda\leq1)$ is the parameter describing
the spacelike straight path $ x^i  = \xi ^i  + \lambda \left( {x -
\xi } \right)^i $, on a fixed time slice. Here, $ \xi $ is a fixed
point (reference point), and there is no essential loss of
generality if we restrict our considerations to $ \xi ^i=0 $. It
immediately follows that the only nontrivial Dirac bracket is
\begin{equation}
\left\{ {A_i^a \left( x \right),\Pi ^{bj} \left( y \right)}
\right\}^ *   = \delta ^{ab} \delta _i^j \delta ^{(2)} \left( {x -y} \right)
- \int\limits_0^1 {d\lambda } \left( {\delta ^{ab} \frac{\partial
}{{\partial x^i }} - gf^{abc} A_i^c \left( x \right)} \right)x^j
\delta ^{(2)} \left( {\lambda x - y} \right). \label{dil45}
\end{equation}

Now, we move on to compute the interaction energy between point-like
sources in the theory under consideration, where a fermion is
localized at the origin $ {\bf 0}$ and an antifermion at $ {\bf y}$.
As already mentioned, to do this we shall calculate the expectation
value of the energy operator, $ H$, in the physical state,
$|\Phi\rangle $. From our above discussion, we see that
$\left\langle H \right\rangle _\Phi$ reads
\begin{equation}
\left\langle H \right\rangle _\Phi   = \frac{1}{2}tr\left\langle
\Phi \right|\int {d^2 x} \left\{ {{\bf \Pi} ^a \left( {1 + \frac{{
m^2 }}{\Delta + a^2 }} \right)^{ - 1} {\bf \Pi} ^a } \right\}\left|
\Phi  \right\rangle
 + \frac{1}{{4
}}tr\left\langle \Phi  \right|\int {d^2 x} F_{ij}^a \left( {1 +
\frac{{m^2 }}{\Delta + a^2 }} \right)F^{aij} \left| \Phi
\right\rangle. \label{dil50}
\end{equation}
At this stage, we recall that the physical state can be written as
\begin{equation}
\left| \Phi  \right\rangle  = \overline \psi  \left( {\bf y}
\right)U\left( {{\bf y},{\bf 0}} \right) \psi \left( {\bf 0}
\right)\left| 0 \right\rangle, \label{dil55}
\end{equation}
where $U\left( {{\bf y},{\bf 0}} \right) \equiv P\exp \left(
{ig\int_{\bf 0}^{\bf y} {dz^i A_i^a \left( z \right)T^a } }
\right)$. As before, the line integral is along a spacelike path on
a fixed time slice, $P$ is the path-ordering prescription and
$\left| 0\right\rangle$ is the physical vacuum state. From the
foregoing Hamiltonian structure, and since the fermions are taken to
be infinitely massive (static), we then get
\begin{equation}
\left\langle H \right\rangle _\Phi   = \left\langle H \right\rangle
_0  + V^{\left( 1 \right)} + V^{\left( 2 \right)},\label{dil60}
\end{equation}
where $\left\langle H \right\rangle _0  = \left\langle 0
\right|H\left| 0 \right\rangle$. The $V^{\left( 1 \right)}$ and
$V^{\left( 2 \right)}$ terms are given by
\begin{equation}
V^{\left( 1 \right)}  = \frac{1}{2}tr\left\langle \Phi \right|\int
{d^2 x} \Pi ^{ai}\frac{{\nabla^2 }}{{\nabla^2 - M^2 }}\Pi
^{ai}\left| \Phi \right\rangle, \label{dil70}
\end{equation}
\begin{equation}
V^{\left( 2 \right)}  =  - \frac{a^2  }{2}tr\left\langle \Phi
\right|\int {d^2 x} \Pi ^{ai}\frac{1}{{\nabla^2  - M^2 }}\Pi
^{ai}\left| \Phi  \right\rangle, \label{dil75}
\end{equation}
with $M^2  \equiv a^2  +  m^2$. From  (\ref{dil45}), one
distinguishes an Abelian part (proportional to $C_{F}$) and a
non-Abelian part (proportional to the combination $C_{F}C_{A}$) for
both $V^{\left( 1 \right)}$ and $V^{\left( 2 \right)}$. After some
lengthy, but straightforward manipulations, we find that, unlike to
the $(3+1)$-dimensional case, the non-Abelian contribution to the
$V^{\left( 2 \right)}$ term is zero. This then implies that, at
leading order in $g$, the $V^{\left( 1 \right)}$ and $V^{\left( 2
\right)}$ terms are essentially Abelian. As a consequence,
(\ref{dil70}) and (\ref{dil75}) take the form
\begin{equation}
V^{\left( 1 \right)}  =  - \frac{{g^2 }}{2}\frac{1}{2} tr\left( {T^a
T^a } \right)\int {d^2 x} \int_{\bf 0}^{\bf y} {dz_i^ {\prime} }
\delta ^{\left( 2 \right)} \left( {x - z^ {\prime}  } \right)
 \frac{{\nabla _x^2 }}{{\nabla _x^2  - M^2 }}\int_{\bf
0}^{\bf y} {dz_i } \delta ^{\left( 2 \right)} \left( {x - z}
\right), \label{dil80}
\end{equation}
\begin{equation}
V^{\left( 2 \right)} = \frac{{g^2 }}{2}\frac{{a^2}}{2} tr\left( {T^a
T^a } \right)\int {d^2 x} \int_{\bf 0}^{\bf y} {dz_i^ {\prime} }
\delta ^{\left( 2 \right)} \left( {x - z^ {\prime}  } \right)
\frac{1}{{\nabla _x^2  - M^2 }}\int_{\bf 0}^{\bf y} {dz^i } \delta
^{\left( 2 \right)} \left( {x - z} \right). \label{dil85}
\end{equation}
According to our earlier procedure \cite{GaeteW}, we find that the
potential for two opposite charges located at ${\bf 0}$ and ${\bf
y}$ becomes
\begin{equation}
V =  - \frac{{g^2 }}{{2\pi }}C_F K_0 (ML) + \frac{{g^2 }}{4}C_F
\frac{{a^2 }}{M}L, \label{dil90}
\end{equation}
where $\left| {\bf y} \right| \equiv L$, $tr(T^{a}T^{a})=C_{F}$ and
$K_0 (ML)$ is a modified Bessel function. Expression (\ref{dil90})
immediately shows that a linearly increasing potential is still
present in the three-dimensional case, corroborating the key role
played by the dilaton field. Interestingly, it is observed that this
is exactly the result obtained for $D=3$ models of antisymmetric
tensor fields that results from the condensation of topological
defects as a consequence of the Julia-Toulouse mechanism
\cite{GaeteW}. In this context, it may be recalled that the
existence of a confining phase for a continuum three-dimensional
Abelian $U(1)$ gauge theory was first found by Polyakov
\cite{Polyakov}, by including the effects due to the compactness of
$U(1)$ group. We further note that for the zero mode case (a=0), the
confining term disappears in (\ref{dil90}). However, from
(\ref{dil25}) we must sum over all the modes in (\ref{dil90}).
Notice that the expression for the coefficient of the linear
potential is given by
\begin{equation}
\sigma  = \frac{{g^2 }}{4}C_F \frac{1}{R}\sum\limits_n {\frac{{n^2
}}{{\sqrt {n^2  + m^2 R^2 } }}}. \label{dil95}
\end{equation}
In the limit $R\to \infty$, we can see that the contributions from
the low $n$ modes are automatically suppressed. Only the higher
modes ($mR\sim n$) are responsible for the finite value of $\sigma$,
namely, $\sigma_0  = \frac{3}{8}C_F g^2 m$. Therefore, according to
expression (\ref{dil90}) the linear piece of the potential stands
and its slope is given by $\sigma_0$.

\section{Related models}

Now, in order to check the consistency of our procedure, it is
instructive to compare our result (\ref{dil90}) with related
three-dimensional models. To do this, we shall begin by recalling
the phenomenological model studied in \cite{GaeteS,GaeteS2}, which
contains both a Yang-Mills and a Born-Infeld term:
\begin{equation}
{\cal L}_{eff}^{\left( {2 + 1} \right)}  =  - \frac{1}{4}F_{\mu \nu
}^a F^{a\mu \nu }  + \frac{M}{2}\sqrt { - F_{\mu \nu }^a F^{a\mu \nu
} }. \label{dil110}
\end{equation}
As was explained in \cite{GaeteS,GaeteS2}, the constant $M$
spontaneously breaks the scale invariance. Recalling again that, by
imposing spherical symmetry, the interaction energy can be exactly
determined. Then, the Lagrangian density (\ref{dil110}) becomes
\begin{equation}
{\cal L}_{eff}^{\left( {2 + 1} \right)}  =  - 2\pi r\left\{
{\frac{1}{{4V}}F_{\mu \nu }^a F^{a\mu \nu }  + \frac{{M^2
}}{4}\frac{V}{{V - 1}}} \right\}, \label{dil115}
\end{equation}
where $V$ is an auxiliary field. Here $\mu,\nu=0,1$, where $x^{1}\equiv
 r = \left| {\bf x} \right|$. We further observe that the
quantization of this theory can be done in a similar manner to that
in \cite{GaeteS,GaeteS2}. This leads to the expectation value
\begin{equation}
\left\langle H \right\rangle _\Phi   = tr\left\langle \Phi
\right|\int {d^2 x} \left( {\frac{{\Pi ^{ai} \Pi ^{ai} }}{{4\pi x}}
+ \frac{{\left| M \right|}}{{\sqrt 2 }}\sqrt {\Pi ^{ai} \Pi ^{ai} }
} \right)\left| \Phi  \right\rangle \nonumber \\
+tr\left\langle \Phi  \right|\int {d^2 x} \frac{1}{4}F_{ij}^a
F^{aij} \left| \Phi  \right\rangle. \label{dil120}
\end{equation}
Once again exploiting the previous procedure leading to
(\ref{dil90}), we find that
\begin{equation}
V = \frac{{g^2 }}{{4\pi }}C_F \ln \left( {\eta L} \right) +
\frac{{\left| M \right|g}}{{\sqrt 2 }}tr\left( {v^{1a} e^1 T^a }
\right)L, \label{dil125}
\end{equation}
where $\eta$ is a massive cutoff and,  $e^{1}$ is a unit vector
starting at ${\bf 0}$ and ending at ${\bf y}$. We further note that,
similarly to the previous case, confinement arose as an Abelian
effect. Mention should be made, at this point, to the results
obtained in Refs. \cite{Schro,Lavelle} for a $(2+1)$-dimensional
$SU(N)$ Yang-Mills theory
\begin{equation}
V(L) = \frac{{g^2 C_F }}{{2\pi }}\log (g^2 L) + \frac{7}{{64\pi }}
g^4 C_F C_A L, \label{dil126}
\end{equation}
where $C_{F}$ and $C_{A}$ are the Casimir group factors. Hence, we
see that the result (\ref{dil125}) agrees with (\ref{dil126}). Let
us mention here that, in order to handle the square root in
expression (\ref{dil120}), we have written  $\Pi ^{ai}  = v^a \Pi
^{ai}$, where $v^{a}$ is a constant vector in color space
\cite{GaeteS2}. In this way, both color and Lorentz symmetries have
been explicitly broken. In order to understand this connection
between confinement and nonconservation of the Lorentz symmetry, we
now examine a three-dimensional Lorentz and CPT violating
Maxwell-Chern-Simons theory \cite{Helayel}. We also point out that
this model was obtained after a reduction to $(2+1)$ dimensions of
an Abelian gauge model with nonconservation of the Lorentz and CPT
symmetries \cite{Jackiw}. Thus, we have
\begin{equation}
{\cal L} =  - \frac{1}{4}F_{\mu \nu }^2  + \frac{1}{2}\partial
_\mu \varphi \partial ^\mu  \varphi  - \frac{1}{2}M_A^2 \varphi ^2
+ \frac{s}{2}\varepsilon _{\mu \nu \lambda } A^\mu  \partial ^\nu
A^\lambda
- \varphi \varepsilon _{\mu \nu \lambda } v^\mu
\partial ^\nu  A^\lambda, \label{dil130}
\end{equation}
where $\varphi$ is a scalar field and $v^\mu$ is a constant vector.
Integrating over $\varphi$, we get
\begin{equation}
{\cal L} =  - \frac{1}{4}F_{\mu \nu }^2  + \frac{1}{8}\overline v
^{\nu \lambda } F_{\nu \lambda } \frac{1}{{\Delta  + M_A^2
}}\overline v ^{\gamma \beta } F_{\gamma \beta }, \label{dil135}
\end{equation}
where we have defined $\overline v ^{\nu \lambda } \equiv
\varepsilon ^{\mu \nu \lambda } v_\mu$. According to our earlier
procedure, the expectation value takes the form
\begin{equation}
\left\langle H \right\rangle _\Phi   = \left\langle \Phi
\right|\int {d^2 x} \left\{ {\frac{1}{2}\Pi ^i \frac{{\nabla ^2
}}{{\nabla ^2  - \overline M _A^2 }}\Pi ^i } \right\}\left| \Phi
\right\rangle
-\frac{{M_A^2 }}{2}\left\langle \Phi  \right|\int {d^2 x} \left\{
{\Pi ^i \frac{1}{{\nabla ^2  - \overline M _A^2 }}\Pi ^i }
\right\}\left| \Phi  \right\rangle, \label{dil140}
\end{equation}
where $\overline M _A^2  \equiv M_A^2  + \tilde v^2$. As a
consequence, the static potential is given by \cite{GaeteW}:
\begin{equation}
V =  - \frac{{g^2 }}{{2\pi }}C_F K_o \left( {\overline M _A L}
\right) + \frac{{g^2 M_A^2 C_F }}{{4\overline M _A }}L.
\label{dil145}
\end{equation}
The above potential profile is analogous to the one encountered in
our previous analysis for gluodynamics in curved space-time
(\ref{dil90}).

\section{Final Remarks}

To conclude, the above connections are of interest from the point of
view of providing unifications among diverse models. More
interestingly, it was shown that our result (\ref{dil90}) agrees
with that of the condensation of topological defects as a
consequence of the Julia-Toulouse mechanism. However, although both
approaches lead to confinement, the above analysis reveals that the
mechanism of obtaining a linear potential is quite different. We
stress here the role played by dilaton in yielding confinement: its
mass contribute linearly to the string tension. We also draw the
attention to the fact that the higher modes are the responsible for
the non-trivial value of the string tension. Our explicit
calculation show that the low $n$ modes are decoupled in the limit
$R$ going to infinity. It would be interesting to employ this
analysis to fit with results coming from lattice calculations.

\section{ACKNOWLEDGMENTS}

One of us (PG) wants to thank the Field Theory Group of the CBPF for
hospitality and PCI/MCT for support. This work was supported in part
by Fondecyt (Chile) grants 1050546 and 7070105. (J.H-N) expresses
his gratitude to CNPq and the staff of the Department of Physics of
the Universidad T\'{e}cnica Federico Santa Mar\'{\i}a for the
pleasant stay.

\end{document}